\documentclass[epj,nopacs,final]{svjour_arxiv}

\usepackage{subfig}
\usepackage{amsmath}
\usepackage{url}
\usepackage{xspace}
\usepackage{verbatim}
\usepackage{graphicx}

\newcommand{\cm}{\ensuremath{\,\text{cm}}\xspace}
\newcommand{\m}{\ensuremath{\,\text{m}}\xspace}
\newcommand{\GeV}{\ensuremath{\,\text{GeV}}\xspace}

\newcommand{\neighbours}{\textsc{Gr\-av\-Net}\xspace}
\newcommand{\aggregators}{\textsc{Gar\-Net}\xspace}
\newcommand{\EdgeConv}{\textsc{Edge\-Conv}\xspace}
\newcommand{\DGCNN}{\textsc{DG\-CNN}\xspace}
\newcommand{\FIN}{\ensuremath{F_{\mathrm{IN}}}\xspace}
\newcommand{\FOUT}{\ensuremath{F_{\mathrm{OUT}}}\xspace}
\newcommand{\FLR}{\ensuremath{F_{\mathrm{LR}}}\xspace}
\newcommand{\FLRtilde}{\ensuremath{\tilde{F}_{\mathrm{LR}}}\xspace}

\newcommand{\figwidth}{0.48}

\title{\boldmath Learning representations of irregular particle-detector geometry with distance-weighted graph networks}

\author{Shah Rukh Qasim\inst{1,2} (14beesqasim@seecs.edu.pk) \and Jan Kieseler\inst{1} (jan.kieseler@cern.ch) \and Yutaro Iiyama\inst{1} (yutaro.iiyama@cern.ch) \and Maurizio Pierini\inst{1} (maurizio.pierini@cern.ch)
}
\institute{CERN, Geneva, Switzerland \and National University of Sciences and Technology, Islamabad, Pakistan}

\authorrunning{S.R. Qasim et al.}
\titlerunning{Distance-weighted graph networks for irregular particle-detector geometries}

\abstract{We explore the use of graph networks to deal with irregular-geometry detectors in the context of particle reconstruction. Thanks to their representation-learning capabilities, graph networks can exploit the full detector granularity, while natively managing the event sparsity and arbitrarily complex detector geometries. We introduce two distance-weighted graph network architectures, dubbed \aggregators and \neighbours layers, and apply them to a typical particle reconstruction task. The performance of the new architectures is evaluated on a data set of simulated particle interactions on a toy model of a highly granular calorimeter, loosely inspired by the endcap calorimeter to be installed in the CMS detector for the High-Luminosity LHC phase. We study the clustering of energy depositions, which is the basis for calorimetric particle reconstruction, and provide a quantitative comparison to alternative approaches. The proposed algorithms provide an interesting alternative to existing methods, offering equally performing or less resource-demanding solutions with less underlying assumptions on the detector geometry and, consequently, the possibility to generalize to other detectors.}

\begin{document}
\maketitle
\flushbottom

\section{Introduction}
\label{sec:intro}

Traditionally, Machine Learning (ML) techniques are a key ingredient to event processing at particle colliders, employed in tasks such as particle reconstruction (clustering), identification (classification), and energy or direction measurement (regression) in calorimeters and tracking devices. The first applications of Neural Networks to High Energy Physics (HEP) date back to the '80s~\cite{Denby:1987rk,Peterson:1988gs,Abreu:1992jp,Denby:1999kv}. Starting with the MiniBooNE experiment~\cite{YANG2005370}, Boosted Decision Trees became the state of the art, and played a crucial role in the discovery of the Higgs boson by the ATLAS and CMS experiments~\cite{Radovic:2018dip}. Recently, a series of studies on different aspects of LHC data taking and data processing workflows have demonstrated the potential of Deep Learning (DL) in collider applications, both as a way to speed up current algorithms and to improve their performance. Nevertheless, the list of DL models actually deployed in the centralized workflows of the LHC experiments remains quite short.\footnote{As an example, at the moment such a list for the CMS experiment consists of a set of b-tagging algorithms~\cite{CMS-DP-2017-013,CMS-DP-2017-027} and a data quality monitoring algorithm for the muon drift tube chambers~\cite{Pol:2018nhq}. Other applications exist at the analysis level, downstream from the centralized event processing. In data analyses, one typically considers abstract four-momenta and not the low-level quantities such as detector hits, making the use of DL techniques easier.} Many of these studies, which are typically proof-of-concept demonstrations, are based on convolutional neural networks~(CNN)~\cite{krizhevsky2012imagenet}, which perform computing vision tasks by applying translation-invariant kernels to {\it raw} digital images. CNN architectures applied on HEP data thus imposes a requirement for the particle detectors to be represented as regular arrays of sensors. This requirement, common to many of the approaches described in Section~\ref{sec:others}, creates problems for realistic applications of CNNs in collider experiments.\footnote{The picture is completely different in other HEP domains. For instance, CNNs have been successfully deployed in neutrino experiments, where the regular-array assumption meets the geometry of a typical detector.}

In this work, we propose novel Deep Learning architectures based on graph networks to improve the performance and reduce the execution time of typical particle-reconstruction tasks, such as cluster reconstruction and particle identification. In contrast to CNNs, graph networks can learn optimal detector-hits representations without making specific assumptions on the detector geometry. In particular, no data preprocessing is required, even for detectors with irregular geometries. We consider the specific case of particle reconstruction in calorimeters, for which this characteristic of graph networks may become especially relevant in the near future. In view of the High-Luminosity LHC phase, the endcap calorimeter of the CMS detector will be replaced by a novel-design digital calorimeter, the High Granularity Calorimeter (HGCAL), consisting of arrays of hexagonal silicon sensor cells interleaved with absorber layers~\cite{HGCAL-TDR}. Being positioned close to the beam pipe and exposed to $\sim 200$ proton-proton collisions on average per bunch crossing, this detector will be characterized by high occupancy over its large number of readout channels. Downstream in the data processing pipeline, the unprecedented number of sensors and their geometry will cause an increase in event size and consequently the computational needs, necessitating novel data processing approaches given the expected computing limitations~\cite{CMS-TP}. The detector we consider in this study, described in detail in Section~\ref{sec:dataset}, is loosely inspired by the HGCAL geometry. In particular, it features a similarly irregular sensor structure, with sensor sizes varying with the detector depth as well as within a single layer. On the other hand, the HGCAL hexagonal sensors were traded for square-shaped sensors, in order to keep the computing resources needed to generate the training data set within a manageable limit. 

As a benchmark application, we consider the basis for all further particle reconstruction tasks in a calorimeter: clustering of the recorded energy deposits into disentangled showers from individual particles. To this purpose, we introduce two novel distance-weighted graph network architectures, the \aggregators and the \neighbours layers, which are designed to provide a good balance between performance and computing resources needs for inference. While our discussion is limited to a calorimetry-related problem, the design of these new layer architectures is such that it automatically generalizes to any kind of sparse data, such as {\it hits} collected by a typical tracking device or reconstructed particle candidates inside a hadronic jet. We believe that architectures of this kind are more practical to deploy in a realistic experimental environment and could become relevant for the LHC experiments, both for offline and real-time event processing and selection as well as shower simulation. 

This paper is structured as follows: Section~\ref{sec:others} reviews related previous works. In Section~\ref{sec:layers}, we describe the \neighbours and \aggregators architectures. Section~\ref{sec:dataset} describes the data set used for this study. Section~\ref{sec:clustering_metrics} introduces the metric used to optimize the networks. Section~\ref{sec:models} describes the models. The results are presented in Sections~\ref{sec:clustering_performance} and \ref{sec:resource_requirements} in terms of accuracy and computational efficiency, respectively. Conclusions are presented in Section~\ref{sec:conclusions}.

\section{Related Work}
\label{sec:others}

In recent years, DL models, and in particular CNNs, have become very popular in different areas of HEP. CNNs were successfully applied to calorimeter-oriented tasks, including particle identification \cite{HGCAL-TDR,carminati2017calorimetry,guest2018deep,deOliveira:2018lqd,FCChh-CDR}, energy regression \cite{HGCAL-TDR,carminati2017calorimetry,deOliveira:2018lqd,FCChh-CDR}, hadronic jet identification \cite{cogan2015jet,komiske2017deep,de2016jet,baldi2016jet}, fast simulation \cite{carminati2017calorimetry,de2017learning,paganini2018calogan,rukhkhattak2018three,Musella:2018rdi} and pileup subtraction in jets~\cite{komiske2017pileup}. Many of these works assume a simplified detector description: the detector is represented as a regular array of sensors expressed as 2D or 3D images, and the problem of overlapping regions at the transition between detector components (e.g. barrel and endcap) is ignored. Sometimes the fixed-grid pixel shape is intended to reflect the typical angular resolution of the detector, which is implicitly assumed to be a constant, while in reality it depends on the energy of the incoming particle.

In order to overcome this practical difficulty with CNN architectures, different HEP-related studies investigating alternative architectures have been performed. In the context of jet identification, several authors studied models based on recurrent~\cite{CMS-DP-2017-013,CMS-DP-2017-027,ATL-PHYS-PUB-2017-003} and recursive~\cite{Louppe:2017ipp} networks, graph networks~\cite{Qu:2019gqs}, and DeepSets~\cite{Komiske:2018cqr}. Recurrent architectures have also been studied for event classification~\cite{Nguyen:2018ugw}. In general, these approaches take as input a particle-based representation of an event and thus are easier to apply in applications running after a global event reconstruction based on a particle-flow algorithm~\cite{Sirunyan:2017ulk,Aaboud:2017aca}.

Outside the HEP domain, overcoming the necessity for a regular structure motivated original research to use graph-based networks~\cite{scarselli2009graph}, which in general are suited for processing point-wise data with no regular structure by representing them as vertices in a graph. A comprehensive overview of various graph-based networks can be found in Ref.~\cite{battaglia2018relational}. In a typical implementation of a graph-based network, the vertices are connected according to some predefined criteria at the preprocessing stage. The connections between the vertices (edges) then define paths of information exchange~\cite{defferrard2016convolutional,velickovic2017graph}. In some cases, the edge and vertex properties are used to infer attention (weight) assigned to each neighbour during this information exchange, while leaving the neighbour relations (adjacency matrix) unchanged~\cite{Selvi2018}.
Some of these architectures have already been considered for collider physics, in the context of jet tagging~\cite{Henrion2017NeuralMP}, event topology classification~\cite{Abdughani:2018wrw}, and for pileup subtraction~\cite{Martinez:2018fwc}.

Particularly interesting for irregular detectors are, however, networks that are capable of learning the geometry, as studied in combination with message passing~\cite{gilmer2017neural}. Within this approach, the adjacency matrix is trainable. 
In other words, the neighbour relations, which encode the detector geometry, are not imposed at the preprocessing stage but are inferred from the input data.
Although this approach is promising, its downside is the need to connect all vertices with each other, which makes it computationally challenging for graphs with a large number of vertices as the memory requirement becomes forbiddingly high. This problem is overcome by defining only a subset of connections between neighbours in a learnable space representation, where the edge properties of each vertex to a limited number of its neighbours are used to calculate a new feature representation per vertex, which is then passed to the next layer of similar structure~\cite{wang2018dynamic}. 
This approach is implemented in the \EdgeConv layer and the corresponding \DGCNN model~\cite{wang2018dynamic}. The neighbours are selected based on the new vertex features, which makes it particularly challenging to create a gradient for training with respect to the neighbour selection. The \DGCNN model works around this issue by using the edge features themselves. However, due to the dynamic calculation of neighbour relations in high-dimensional space, this network still requires substantial computing resources, which would make its use for triggering purposes in collider detectors unfeasible.


\section{The \neighbours and \aggregators layers}
\label{sec:layers}

\begin{figure*} 
\centering
\includegraphics[width=0.7\textwidth]{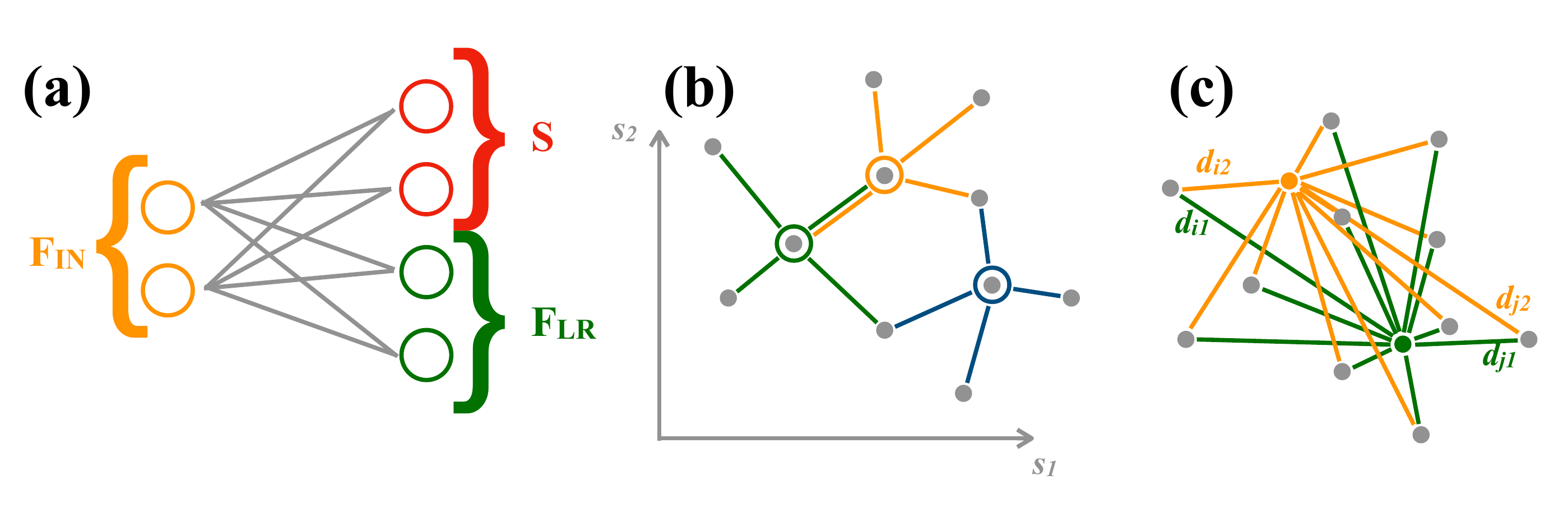}
\includegraphics[width=0.7\textwidth]{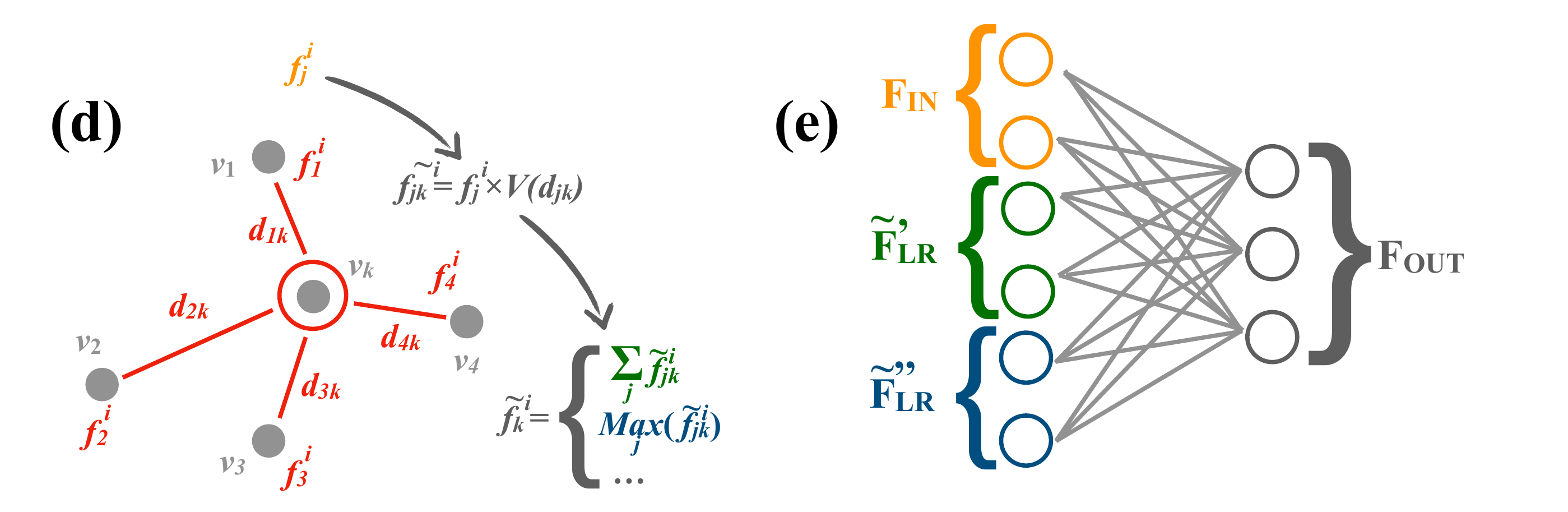}
\caption{Pictorial representation of the data flow across the \aggregators and the \neighbours layers. (a) The input features \FIN
of each $v_i \in V$ are processed by a dense neural network with two output arrays: a set of learned features \FLR and spatial information $S$ in some learned representation space. (b) In the case of the \neighbours layer, the $S$ quantities are interpreted as the coordinates of the vertices in some abstract space. The graph is built in this space, connecting each $v_i$ to its $N$ closest neighbors (N=4 in the figure), using the euclidean distance $d_{ij}$ between the vertices to rank the neighbors. (c) In the case of the \aggregators layer,  the $S$ quantities are interpreted as the distances between the
vertices and a set of $S$ aggregators in some abstract space. The graph is then built connecting each $v_i$ vertex to each $a_j$ aggregator, and the $S$ quantities are the $d_{ij}$ euclidean distances. (d) Once the graph structure is established, the $f_j^i$ features of the $v_j$ vertices connected to a given vertex or aggregator $v_k$ are converted into the ${\tilde f}_{jk}^i$ quantities, through a potential (function of $d_{jk}$). The corresponding information is then gathered across the graph and turned into a new feature ${\tilde f}_{k}^i$ of $v_k$ (e.g. summing over the edges, or taking the maximum). (e) For each choice of gathering function, a new set of features ${\tilde f}_{k}^i \in \FLRtilde$ is generated. The \FLRtilde vector is concatenated to the initial \FIN vector. The resulting feature vector is given as input to a dense neural network with $\tanh$ activation, which returns the output representation \FOUT.
\label{fig:graph_cartoon}}
\end{figure*}

The neural network layers proposed in this study are designed to provide competitive performance on particle reconstruction tasks while dealing with data sparsity in an efficient way. These architectures aim to keep a trainable space representation at minimal computational costs. 
The layers receive as input a $B\times V \times \FIN$ data set, consisting of a batch of $B$ examples, each represented by a set of $V$ detector hits, embedded in the network set through \FIN features. For instance, the \FIN features could include the Cartesian coordinates of a given sensor, its address (layer number, module number, etc.), the sensor time stamp, the recorded energy, etc. 

A pictorial representation of the operations performed by the two layers is shown in Fig.~\ref{fig:graph_cartoon}. For both architectures, the first step is to apply a dense\footnote{Here and in the following, {\it dense} layer refers to a learnable weight-matrix multiplication and bias vector addition with respect to the last feature dimension, with shared weights over all other dimensions. In this case, the weights and bias are applied to the vertex features \FIN and shared over the vertices $V$. This can also be thought of as a 2D convolution with a $1 \times 1$ kernel.} neural network to each of the $V$ detector hits, deriving from the \FIN features two output arrays: the first array ($S$) is interpreted as a set of coordinates in some learned representation space (for the \neighbours layer) or as the distance between the considered vertex and a set of $S$ aggregators (for the \aggregators layer); the second array (\FLR) is interpreted as a learned representation of the vertex features. At this point, a given input example of initial dimension $V \times \FIN$ is converted into a graph with $V$ vertices in the abstract space identified by $S$. Each vertex is represented by the \FLR features, derived from the initial inputs. The projection from the $V \times \FIN$ to this graph is linear, with trainable weights and bias vectors.

The main difference between the \neighbours and the \aggregators architectures is in the way the $V$ vertices are connected when building the graph. In the case of the \neighbours layer, the Euclidean distances $d_{jk}$ between  $(j, k)$ pairs of vertices in the $S$ space are used to associate to each vertex its closest $N$ neighbors. In the case of the \aggregators layer, the graph is built connecting each of the $V$ vertices to a set of $\text{dim}(S)$ aggregators. What is learned by $S$, in this case, is the distance between a vertex and each of the aggregators.

Once the edges of the graph are built, each vertex (aggregator) of the  \neighbours{ }(\aggregators) layer collects the information associated with the \FLR features across its edges. This is done in three steps: 
\begin{enumerate}
\item The quantities 
\begin{equation}
\label{eq:gauss}
\tilde{f}^i_{jk} = f^i_j \times V(d_{jk})
\end{equation}
are computed for the feature $f^i$ of each of the vertices $v_j$ connected to a given vertex or aggregator $v_k$, scaling the original value by a {\it potential}, function of the euclidean distance $d_{jk}$, giving the {\it gravitational network} \neighbours its name.
The potential function is introduced to enhance the contribution of close-by vertices. For this reason, $V$ has to be a decreasing function of $d_{jk}$. In this study, we use a Gaussian potential $V(d_{jk}) = \exp{(-d_{jk}^2)}$ for the \neighbours layer\footnote{A gravitational potential ($-1/d$) has singularities at $d=0$ and therefore cannot be used, however the potential we are using has a similar qualitative effect of pulling together vertices.} and an exponential potential $V(d_{jk}) = \exp{(-|d_{jk}|)}$ for the \aggregators layer.
\item The $\tilde{f}^i_{jk}$ functions computed from all the edges associated to a vertex of aggregator $v_k$ are combined, generating a new feature $\tilde{f}^i_{k}$ of $v_k$. For instance, we consider the average of the $\tilde{f}^i_{jk}$ across the $j$ edges and their maximum. 
In our case, it was particularly crucial to extend the choice of aggregator functions beyond the maximum, which was already explored for similar architectures~\cite{wang2018dynamic}. In fact, the mean function (as any other similar function) helped improve the convergence of the model, by taking into account the contribution of all the vertices.
\item Each adopted combination rule in the previous step generates a new set of features \FLRtilde. All of them are concatenated to the original \FIN vector. This extended vector is transformed into a set of \FOUT new vertex features, using a fully connected dense layer with $\tanh$ activation. The concatenation is done for each initial vertex. In the case of the \aggregators layer, this requires an additional step of passing the $\tilde{f}^i_{k}$ features of the $v_k$ aggregators back to the initial vertices, weighted by the $V(d_{jk})$ potential. This information exchange of the {\it garnered} information through the aggregators defines the \aggregators name.
\end{enumerate}
The full process transforms the initial $B\times V \times \FIN$ data set into a $B\times V \times \FOUT$ data set. As common with graph networks, the main advantage comes from the fact that the
\FOUT output (unlike the \FIN input) carries collective information from each vertex and its surrounding, providing a more informative input to downstream processing. Thanks to the distinction between learned space information $S$ and learned features \FLR, the dimensionality of connections in the graph is kept under control, resulting in a smaller memory consumption than, for instance, the  \EdgeConv layer.

The two layer architectures and the models based on them, described in the following sections, are implemented in TensorFlow~\cite{tensorflow2015-whitepaper}.~\footnote{The code for the models and layers can be found in \url{https://github.com/jkiesele/caloGraphNN}}

\begin{figure}[hbtp]
    \centering
    \includegraphics[width=\figwidth\textwidth]{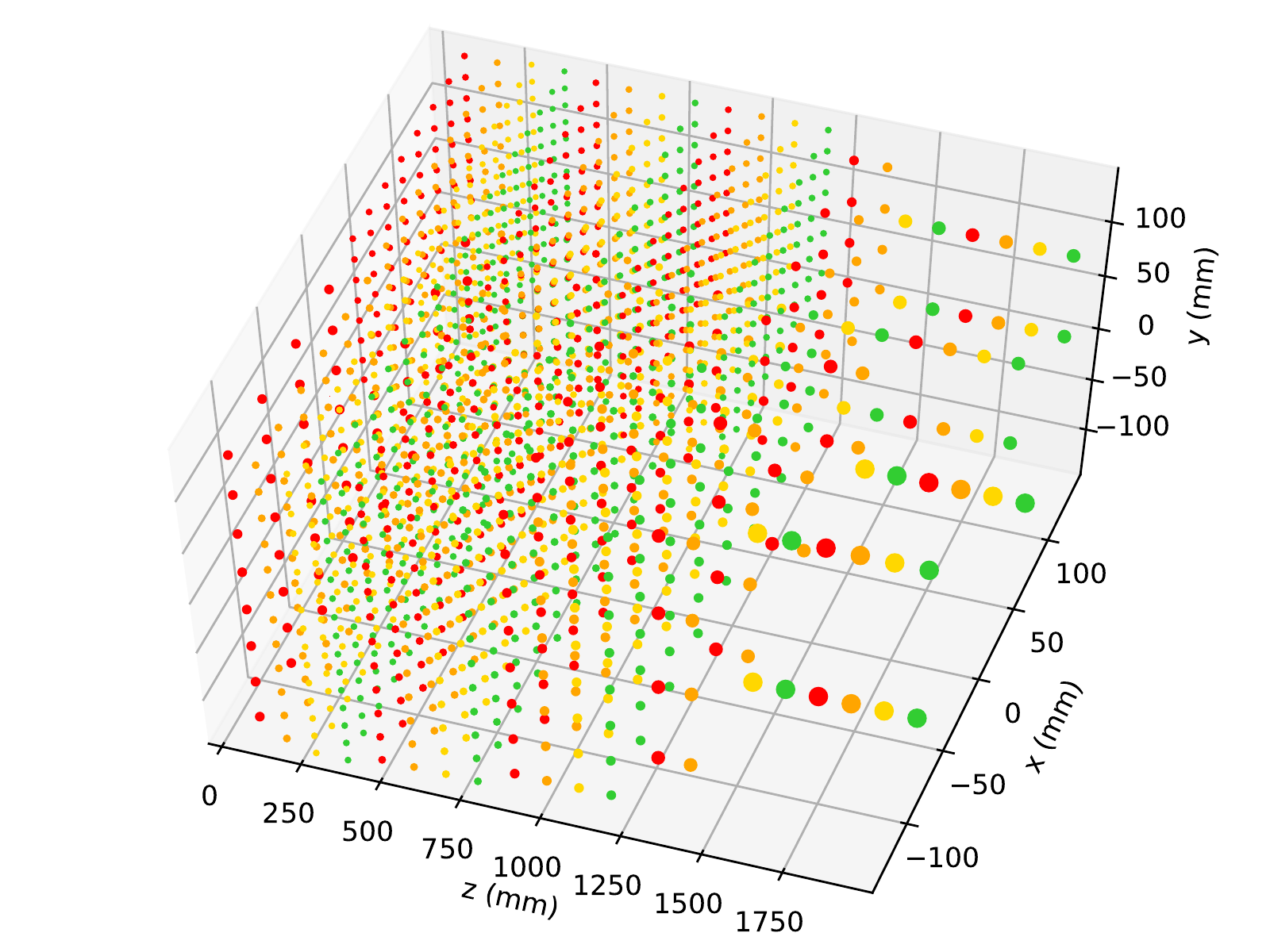}
    \caption{Calorimeter geometry. The markers indicate the centre of the sensors, their size the sensor size. Layers are colour-coded for better visualisation.}
    \label{fig:geometry}
\end{figure}

\section{Data set}
\label{sec:dataset}

The data set used in this paper is based on a simplified calorimeter with irregular geometry, built in GEANT4~\cite{agostinelli2003geant4}. The calorimeter is made entirely of Tungsten, with a width of $30\cm \times 30\cm$ in the x and y directions and a length of 2\m in the longitudinal direction (z), which corresponds to 20 nuclear interaction lengths. The longitudinal dimension is further split into 20 layers of equal thickness. Each layer contains square sensor cells, with a fine segmentation in the quadrant with $x>0$ and $y>0$ and a lower granularity elsewhere. The total number of cells and their individual sizes vary by layer, replicating the basic features of a slightly irregular calorimeter. For more details, see Fig.~\ref{fig:geometry} and Table~\ref{tab:geometry}.

\begin{table}[htbp]
    \centering
    \begin{tabular}{l|c|c}
           Layer & Cells ($x>0$, $y>0$) & Cells elsewhere \\ \hline
        0   & 64 & 48  \\
        1   & 64 & 108  \\
        2--3   & 100 & 192  \\
        4--7   & 64 & 108  \\
        8--11   & 64 & 48  \\
        12--13   & 16 & 12  \\
        14--19   & 4 & 3  \\
    \end{tabular}
    \caption{Number of cells in the finely segmented quadrant and the rest of the layer, for the benchmark calorimeter geometry described in the text.}
    \label{tab:geometry}
\end{table}

Charged pions are generated at $z=-2\m$; the $x$ and $y$ coordinates of the generation vertex are randomly sampled within $|x|<5\cm$ and $|y|<5\cm$. 
The $x$ and $y$ components of the particle momentum are set to 0, while the $z$ component is sampled uniformly between 10 and 100\GeV. The particles therefore impinge the calorimeter front face perpendicularly and shower along the longitudinal direction.

The resulting total energy deposit in each cell, as well as the cell position, width, and layer number, are recorded for each event. These quantities correspond to the \FIN feature vector given as input to the graph models (see Section~\ref{sec:layers}). Each example consists of the result of two overlapping showers. Cell by cell, the energy of two showers is summed and the fraction belonging to each of the showers in each cell is defined as the ground truth. In addition, the position of the largest energy deposit per shower is recorded. If this position is the same for the two overlapping showers, they are considered not separable and the event is discarded. This applies to about 5\% of the events.

In total 16$\,$000$\,$000 events are generated. Out of these, 100$\,$000 are used for validation and 250$\,$000 for testing. The rest is used for training.

\section{Clustering metrics}
\label{sec:clustering_metrics}

To identify individual showers and use their properties, e.g. for a subsequent particle identification task, the energy deposits should be clustered so that overlapping parts are identified without removing important parts of the original shower. Therefore, the clustering algorithms should predict the energy fraction of each sensor belonging to each shower. Lower energy deposits are slightly less important.
These considerations define the loss function:
\begin{equation}
\label{eq:loss}
    L = \sum_k \frac{\sum_i \sqrt{E_{i}t_{ik}} (p_{ik}-t_{ik})^2}{\sum_i \sqrt{E_{i}t_{ik}}}\text{,}
\end{equation}
where $p_{ik}$ and $t_{ik}$ are the predicted and true energy fractions in sensor $i$ and shower $k$. These are weighted by the square root of $E_{i}t_{i}$, which is the total energy deposit in sensor $i$ belonging to shower $k$, to introduce a mild energy scaling within each shower.

In addition, in each event we randomly label one of the showers as the \textit{test} shower and the other as the \textit{noise} shower, and define the clustering energy response $R_k$ of shower $k$ ($k=\text{test, noise}$) as:
\begin{equation}
\label{eq:response}
    R_k = \frac{\sum_i E_{i} p_{ik}}{\sum_i E_{i} t_{ik}}
\end{equation}

\section{Models}
\label{sec:models}

The models need to incorporate neural network layers to identify localized structures as well as to perform information exchange globally between the sensors. This can be achieved either by multiple message passing iterations between neighbouring sensors or a direct global information exchange. Here, we employ a combination of both.
The input to all models is an array of sensors, each holding its recorded energy deposits, global position coordinates, sensor size, and layer number. We compare three different graph-network approaches to a CNN based approach (Binning), presented as a baseline. Each model is designed to contain approximately $100\, 000$ free parameters. The model structure is as follows: 

\begin{itemize}
    \item \textbf{Binning:} a regular grid of $20 \times 20  \times 20$ pixels is imposed on the irregular geometry. Each pixel contains the information of at most one sensor\footnote{Alternative configurations with more than one sensor per pixel were also investigated and showed similar performance.}. The information is concatenated to the mean of these features in all pixels, pre-processed by one $1\times 1 \times 1$ CNN layer with 20 nodes, and then fed through eight blocks of CNN layers. Each block consists of a CNN layer with a kernel of $7\times 7 \times 1$ followed by a layer with a kernel of $1\times 1 \times 3$, each containing 14 filters.  The output of each block is passed to the next block and simultaneously added to a list of all block outputs. All CNN layers employ tanh activation functions.
    Finally, the full list of block outputs per pixel is reshaped to represent the vertices of the graph and fed through a dense layer with 128 nodes and ReLU activation. Different CNN models have also been tested and showed similar or worse performance.
    \item \textbf{\DGCNN model:} adapting the model proposed in Ref~\cite{wang2018dynamic} to our problem, the sensor features are interpreted as positions of points in a 16-dimensional space and fed through one global space transformation followed by four blocks comprising one \EdgeConv layer. Our \EdgeConv layer has a similar configuration as in Ref.~\cite{wang2018dynamic}, with 40 neighbouring vertices and three internal dense layers with ReLu activation acting on the edges with 64 nodes each. The output of the \EdgeConv layer is concatenated with its mean over all vertices and fed to one dense layer with 64 nodes and ReLu activation which concludes the block. The output of each block is passed to the next block and simultaneously added to a list of all block outputs per vertex together with the mean over vertices. This list is finally fed to a dense layer with 32 nodes and ReLU activation.
    \item \textbf{\neighbours model:} the model consists of four blocks. Each block starts with concatenating the mean of the vertex features to the vertex features, three dense layers with 64 nodes and $\tanh$ activation, and one \neighbours layer with $S=4$ coordinate dimensions, $\FLR=22$ features to propagate, and $\FOUT=48$ output nodes per vertex. For each vertex, 40 neighbours are considered. The output of each block is passed as input to the next block and added to a list containing the output of all  blocks. This determines the full vector of vertex features passed to a final dense layer with 128 nodes and ReLU activation.
    \item \textbf{\aggregators model:} The original vertex features are concatenated with the mean of the vertex features and then passed on to one dense layer with 32 nodes and tanh activation before entering 11 subsequent \aggregators layers. These layers contain $S=4$ aggregators, to which $\FLR=20$ features are passed, and $\FOUT=32$ output nodes. The output of each layer is passed to the next and added to a vector containing the concatenated outputs of each \aggregators layer. The latter is finally passed to a dense layer with 48 nodes and ReLU activation. 
\end{itemize}

In all cases, each output vertex of these model building blocks is fed through one dense layer with ReLU activation and three nodes, followed by a dense layer with two output nodes and softmax activation. This last processing step determines the energy fraction belonging to each shower. 
Batch normalisation~\cite{ioffe2015batch} is applied in all models to the input and after each block.

All models are trained on the full training data set using the Adam optimizer~\cite{kingma2014adam} and an initial learning rate of about $3 \times 10^{-4}$, the exact value depending on the model. The learning rate is reduced exponentially in steps to the minimum of $3 \times 10^{-6}$ after 2 million iterations. Once the learning rate has reached the minimum level, it is modulated by 10\% at a fixed frequency, following the method proposed in Ref.~\cite{DBLP:journals/corr/abs-1708-07120}. 

\section{Clustering performance}
\label{sec:clustering_performance}

\begin{figure}[htbp]
    \centering
    \subfloat[Truth]{{\includegraphics[width=\figwidth\textwidth]{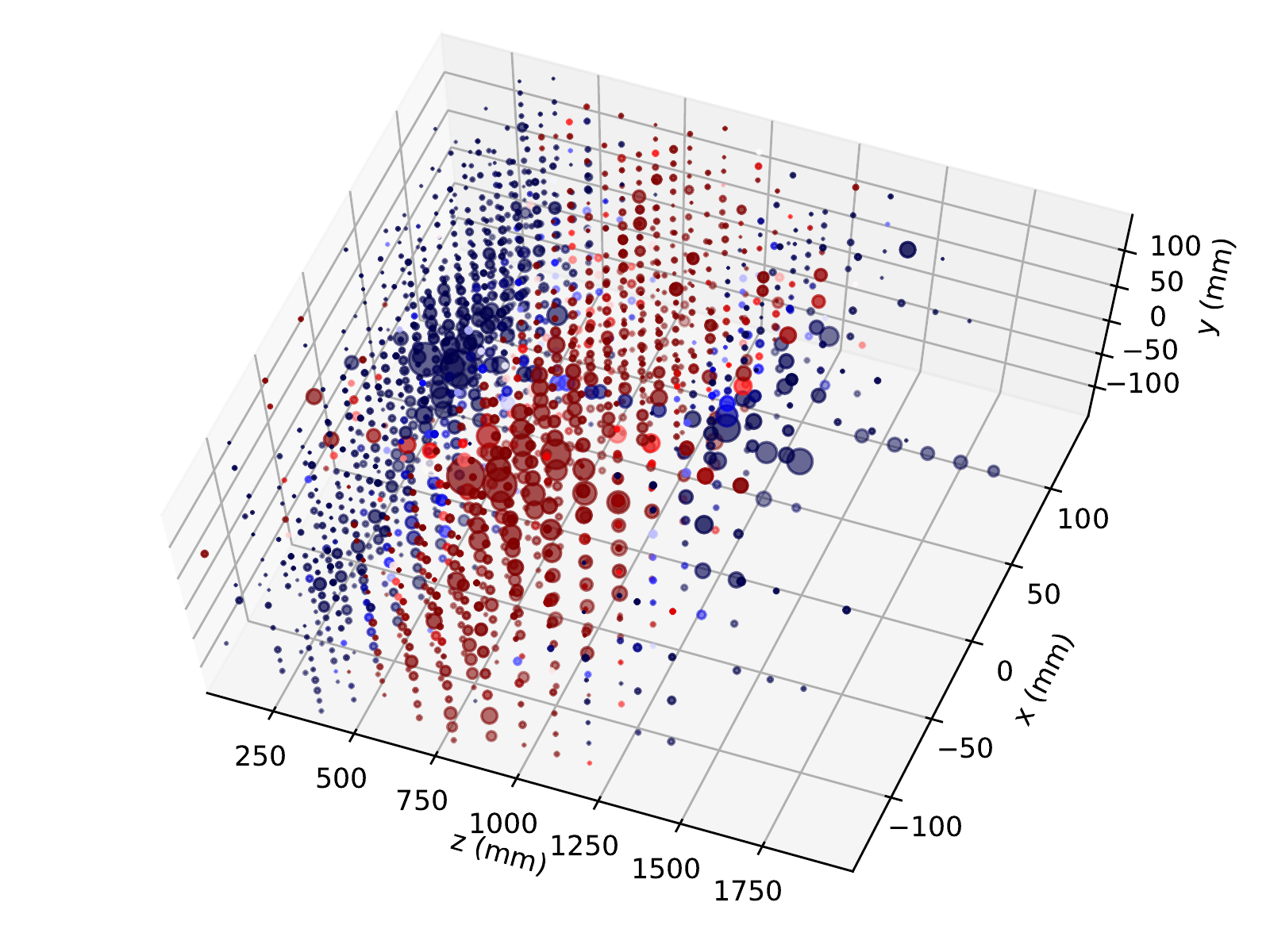} }}%
    \qquad
    \subfloat[Reconstructed]{{\includegraphics[width=\figwidth\textwidth]{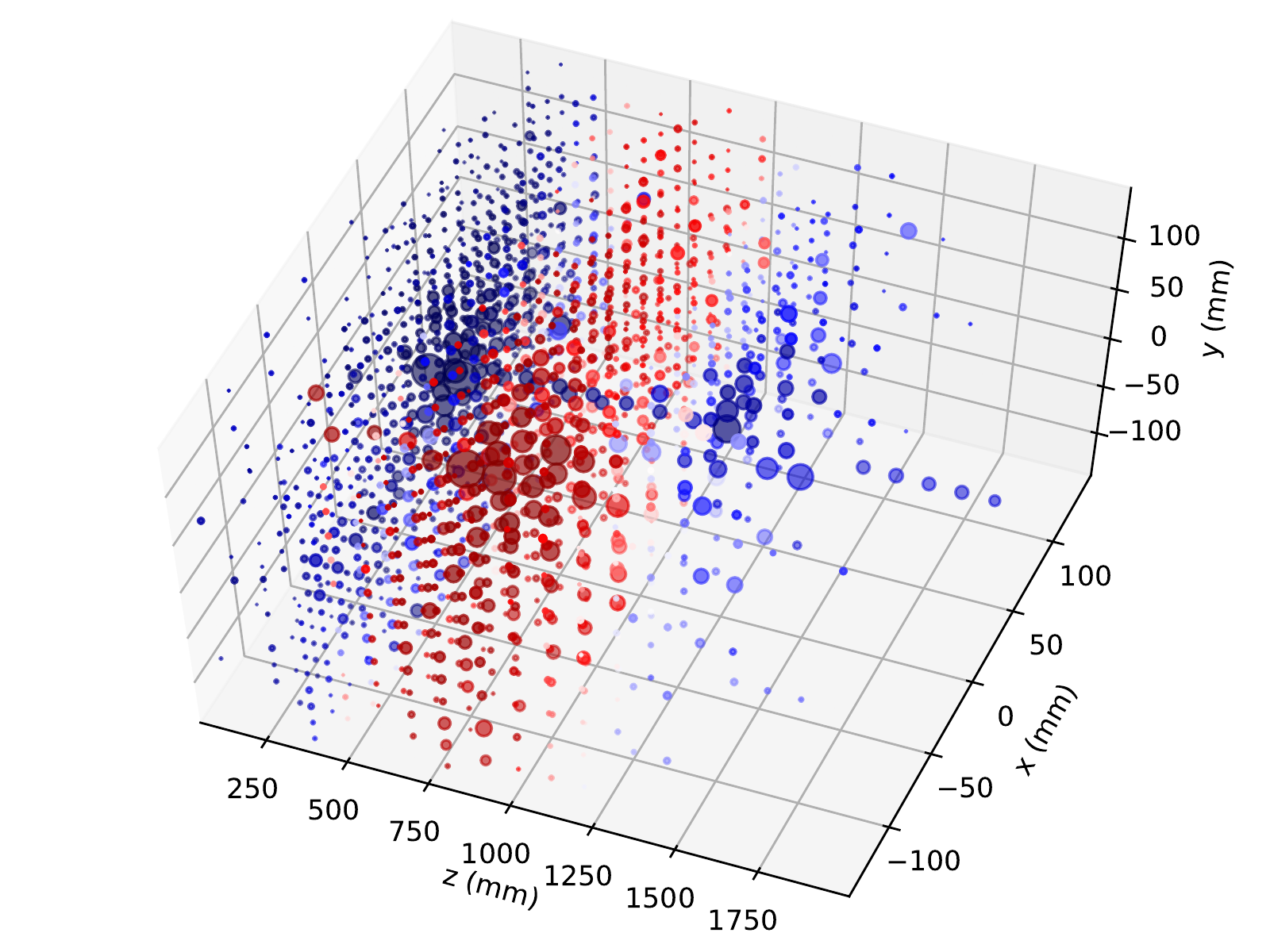} }}%
    \caption{Comparison of true energy fractions and  energy fractions reconstructed by the \neighbours model for two charged pions with an energy of approximately 50\GeV showering in different parts of the calorimeter. Colours indicate the fraction belonging to each of the showers. The size of the markers scales with the square root of the energy deposit in each sensor.}
    \label{fig:example}%
\end{figure}

All approaches described in Section~\ref{sec:models} perform well for clustering purposes. An example is shown in Fig.~\ref{fig:example}, where two charged pions with an energy of approximately 50\GeV enter the calorimeter. One pion loses a significant fraction of energy in an electromagnetic shower in the first calorimeter layers. The remaining energy is carried by a single particle passing the central part of the calorimeter before showering.
The second pion passes the first layers as a minimally ionizing particle and showers in the central part of the calorimeter. Even though the two showers largely overlap, the \neighbours network (shown here as an example) is able to identify and separate the two showers very well. The track within the calorimeter is well identified and reconstructed and the energy fractions properly assigned, even in the parts where the two showers heavily overlap. Similar performance can be observed with the other investigated methods.

Quantitatively, the models are compared with respect to multiple performance metrics. The first two are the mean and the variance of the loss function value ($\mu_L$ and $\sigma_L$) computed according to Equation~\eqref{eq:loss} over the test events. The mean and the variance of the test shower response ($\mu_R$ and $\sigma_R$), where the response is defined in Equation~\eqref{eq:response}, are also compared. While the test shower response follows an approximately normal distribution over majority of the test events, a small outlier population, where the shower clustering fails, are seen to lead $\mu_R$ and $\sigma_R$ to misparametrize the core of the distribution. Therefore, response kernel mean $\mu^*_R$ and variance $\sigma^*_R$, restricted to test showers with response between 0.2 and 2.8, are added to the set of evaluation metrics. In addition, we also compare the clustering accuracy ($A$), defined as the fraction of showers with response between 0.7 and 1.3. Finally, the above set of metrics is duplicated, with the second set using only the sensors with energy fractions between 0.2 and 0.8 in the computation of the loss function and the response. The second set of metrics characterizes the performance of the models in particularly challenging case of reconstructing significantly overlapping clusters. The two sets of metrics are called \textit{inclusive} and \textit{overlap-specific} in the remainder of the discussion.

The metric values are listed in Table~\ref{tab:perf}. Comparing the inclusive metrics, it can be seen that the \neighbours layer outperforms the other approaches, including even the more resource-intensive \DGCNN model. The \aggregators model performance is in between the \DGCNN model and the binning approach in terms of reconstruction of individual shower hit fractions, parametrized by $\mu_{L}$ and $\sigma_{L}$. However, in characteristics related to clustering response, the binning model outperforms the \aggregators and \DGCNN model slightly. On the other hand, with respect to overlap-specific metrics, the graph based approaches outperform the binning approach. The \DGCNN and \neighbours model perform equally well, and the \aggregators model lies in between the binning approach and \neighbours.

\begin{table}[htbp]
\setlength\tabcolsep{3pt}
    \centering
    \caption{Mean and variance of loss, response, and response within the Gaussian kernel as well as clustering accuracy.}
    \begin{tabular}{l|c|c|c|c|c|c|c}
          \multicolumn{8}{c}{Inclusive} \\ \hline
          & $\mu_L$ & $\sigma_L$ &$\mu_R$ & $\sigma_R$  &  $\mu^*_R$ &$\sigma^*_R$ & $A$ \\ \hline
          Binning &0.191&0.017&1.083 &0.183 & 1.046&0.057 & 0.867\\
          \DGCNN &0.174&0.012&1.082 &0.179 & 1.045&0.052 & 0.881\\
          \aggregators &0.182&0.011&1.086 &0.190 & 1.048&0.055 & 0.872\\
          \neighbours &0.172&0.012&1.077 &0.173 & 1.042&0.049 & 0.886\\ \hline \hline
          \multicolumn{8}{c}{Overlap-specific} \\ \hline
          & $\mu_L$ & $\sigma_L$ &$\mu_R$ & $\sigma_R$  &  $\mu^*_R$ &$\sigma^*_R$ & $A$ \\ \hline
          Binning &0.163&0.0045&1.005 &0.099 & 1.004&0.096 & 0.697\\
          \DGCNN &0.154&0.0046&1.004 &0.090 & 1.002&0.087 & 0.728\\
          \aggregators &0.157&0.0048&1.005 &0.095 & 1.004&0.092 & 0.714\\
          \neighbours &0.156&0.0047&1.004 &0.091 & 1.003&0.088 & 0.721\\ \hline
    \end{tabular}
    \label{tab:perf}
\end{table}

One should notice that part of the incorrectly predicted events are actually correctly clustered events in which the test shower is labelled as noise shower (shower swapping). Since the labelling is irrelevant in a clustering problem, this behavior is not a real inefficiency of the algorithm. We denote by $s$ the fraction of events where this behaviour is observed. In Table~\ref{tab:perf_swap}, we calculate the loss for both choices and evaluate the performance parameters for the assignment that minimizes the loss.
The binning model shows the largest fraction of swapped showers. The difference in response between the best-performing \neighbours model and the \aggregators model is enhanced, while the difference between the \neighbours and \DGCNN model scales similarly, likely because of their similar general structure. 


\begin{table}[htbp]
\setlength\tabcolsep{2pt}
    \centering
    \caption{Mean and variance of loss, response, and response within the Gaussian kernel as well as clustering accuracy corrected for shower swapping. The last column shows the fraction of swapped showers.}
    \begin{tabular}{l|c|c|c|c|c|c|c|c}
          \multicolumn{9}{c}{Inclusive} \\ \hline
          & $\mu_L$  & $\sigma_L$  & $\mu_R$ & $\sigma_R$  & $\mu^*_R$ &$\sigma^*_R$ & $A$ & $s\, [\%]$\\ \hline
          Binning    &0.179&0.007&1.076 &0.139 & 1.047&0.054 & 0.875 & 3.2\\
          \DGCNN      &0.167&0.006&1.076 &0.138 & 1.047&0.050 & 0.887 & 2.6\\
          \aggregators &0.176&0.006&1.081 &0.149 & 1.049&0.054 & 0.877 & 2.5\\
          \neighbours    &0.164&0.006&1.071 &0.126 & 1.044&0.047 & 0.892 & 2.7\\ \hline \hline
          \multicolumn{9}{c}{Overlap-specific} \\ \hline
          & $\mu_L$  & $\sigma_L$  & $\mu_R$ & $\sigma_R$  & $\mu^*_R$ &$\sigma^*_R$ & $A$ & $s\, [\%]$\\ \hline
          Binning    &0.160&0.0037&1.005 &0.098 & 1.004&0.095 & 0.699 & 3.2\\
          \DGCNN      &0.152&0.0038&1.003 &0.089 & 1.002&0.086 & 0.729 & 2.6\\
          \aggregators &0.154&0.0040&1.005 &0.094 & 1.003&0.091 & 0.715 & 2.5\\
          \neighbours    &0.152&0.0039&1.004 &0.090 & 1.003&0.087 & 0.722 & 2.7\\ \hline
    \end{tabular}
    \label{tab:perf_swap}
\end{table}

\begin{figure*}[htbp]
    \centering
    \subfloat[Mean]{{\includegraphics[width=\figwidth\textwidth]{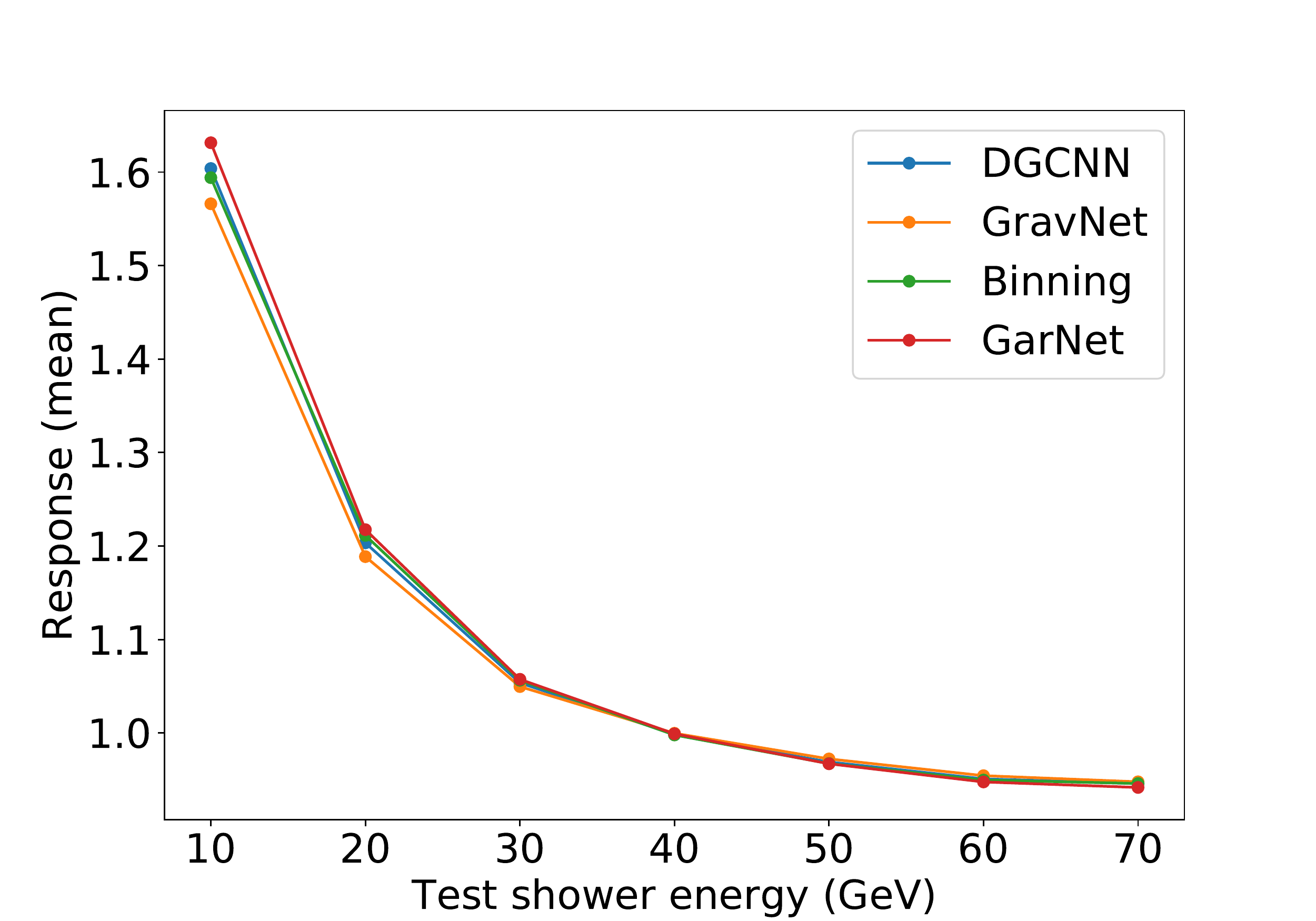}}}%
    \qquad
    \subfloat[Variance]{{\includegraphics[width=\figwidth\textwidth]{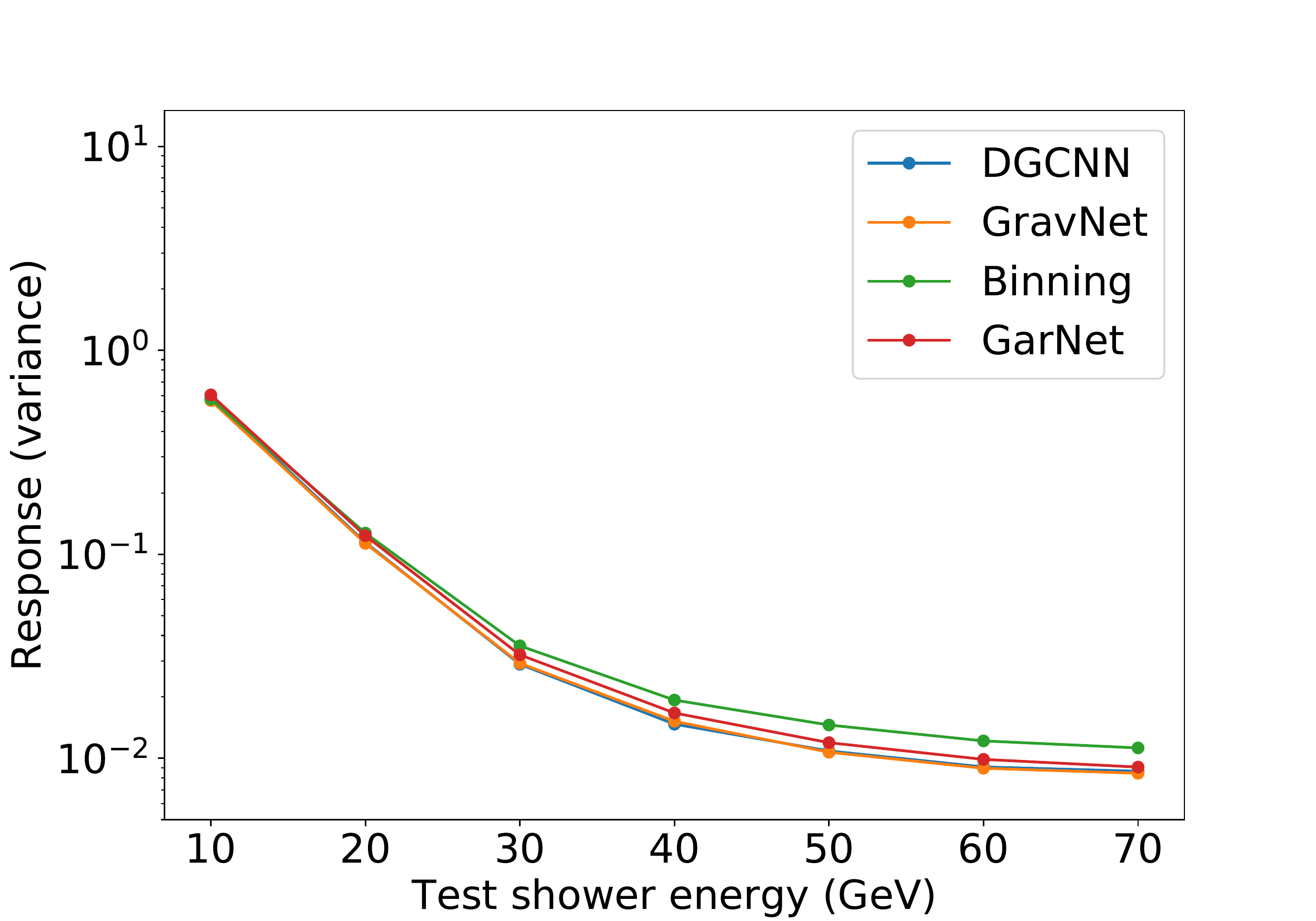}}}
    \\
    \subfloat[Mean]{{\includegraphics[width=\figwidth\textwidth]{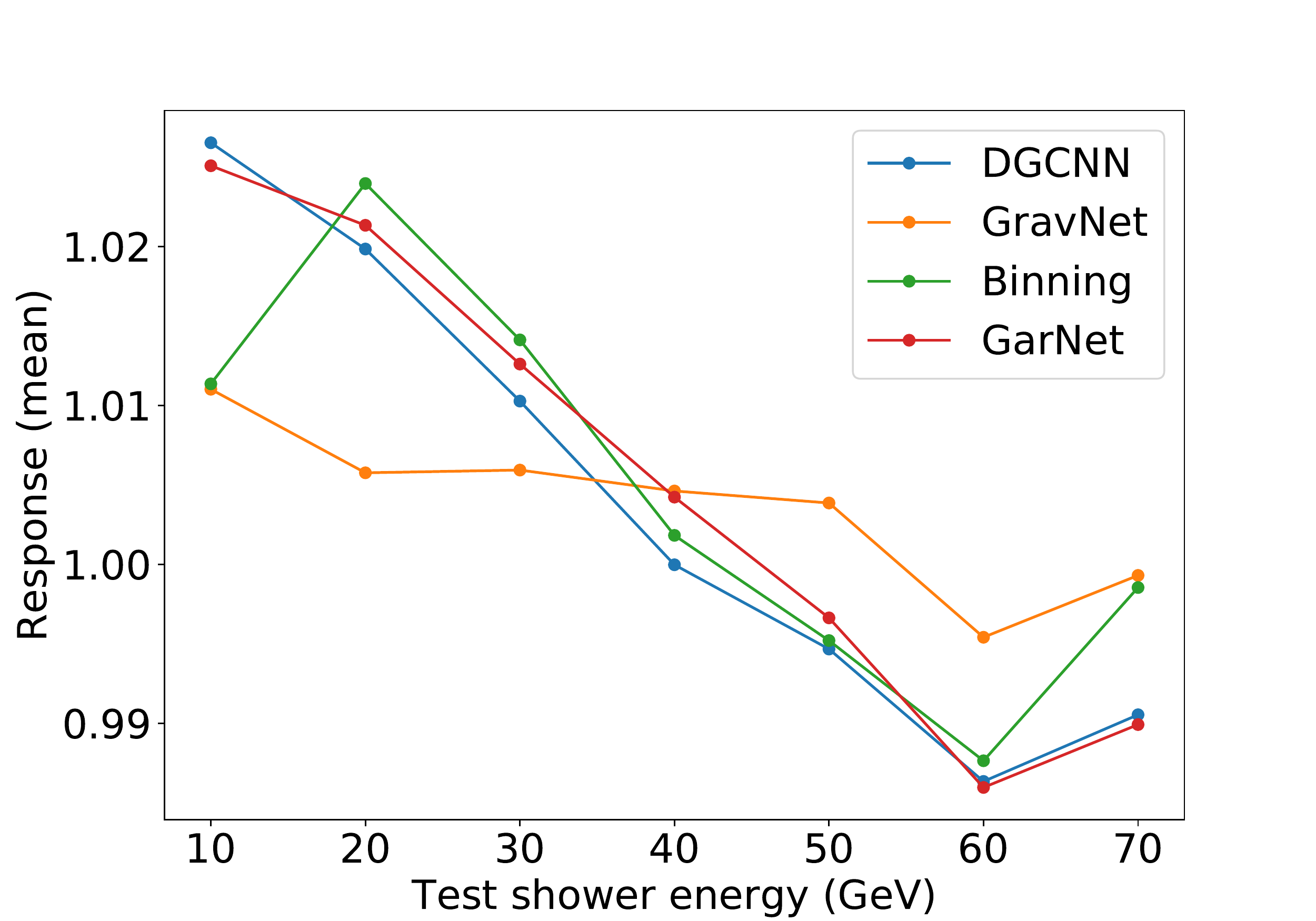}}}%
    \qquad
    \subfloat[Variance]{{\includegraphics[width=\figwidth\textwidth]{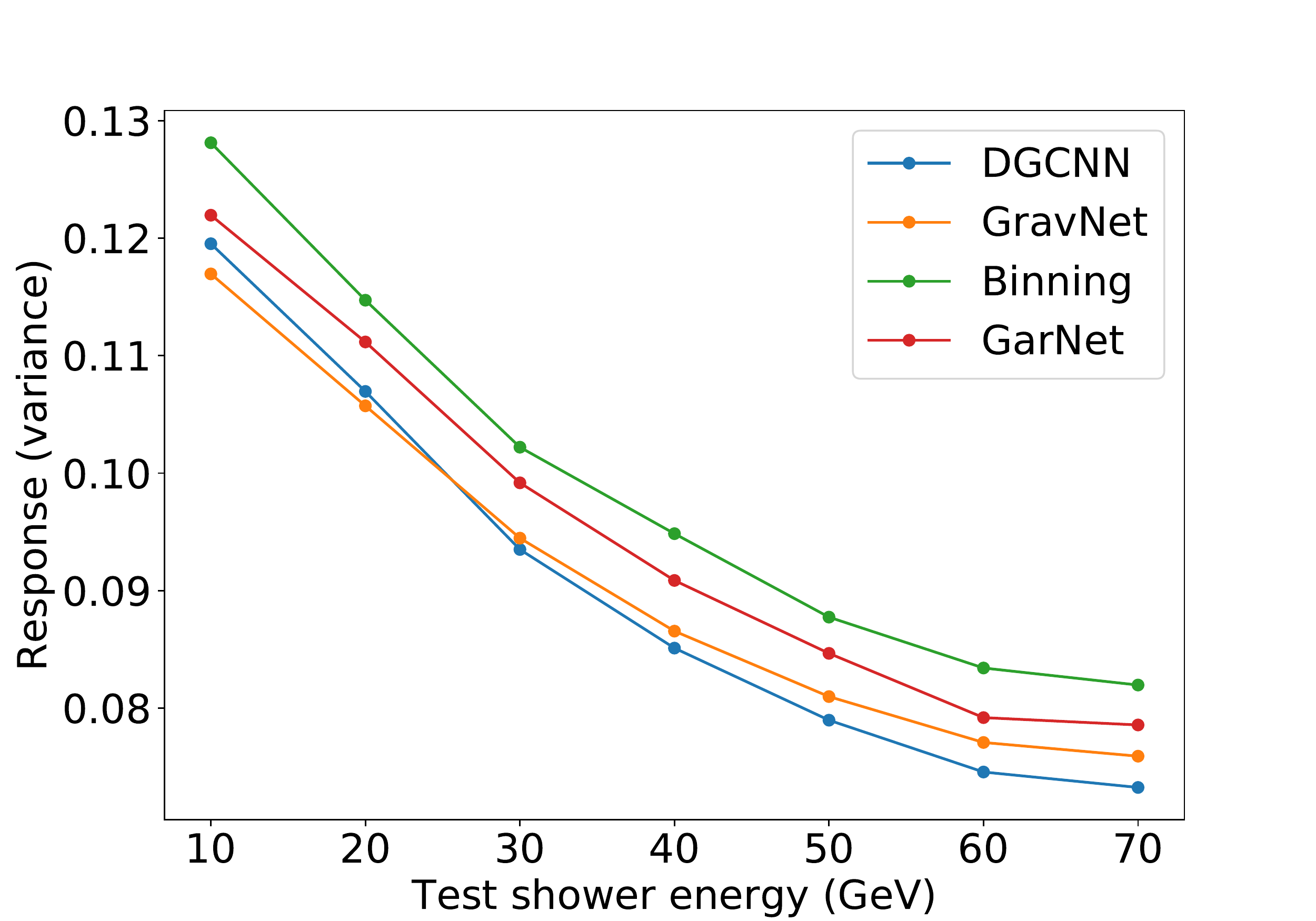}}}%
    \caption{Mean (left) and variance (right) of the test shower response as a function of the test shower energy for full shower (top) and for overlapping shower (bottom), computed summing the true deposited energy. Swapping of the showers is allowed here.}
    \label{fig:mean_response_curve}
\end{figure*}

In Fig.~\ref{fig:mean_response_curve}, the performance of the models are compared in bins of the test shower energy with respect to inclusive and overlap-specific $\mu_R$ and $\sigma_R$. For the inclusive metrics, the \neighbours model outperforms the other models in the full range, and the \aggregators model shows the worst performance, albeit in a comparable range. The resource-intensive \DGCNN model lies in between \neighbours and \aggregators. 

The overall upward bias in the response for lower shower energies warrants an explanation. This bias is a result of edge effects, induced by our choice of using an adapted mean-square error loss to predict a quantity bounded in [0,1] (the energy fraction). This choice of loss function creates an expectation value larger than 0 at a peak value of 0 (and vice-versa at a fraction of 1), and therefore pushes the prediction away from being exactly 0 or 1, leading to an underestimation at high energies and an overestimation at low energies. The design of a customized loss function that eliminates this bias is left to future studies. For the moment, we are interested in a performance comparison between models, all affected by this bias.

For overlap-specific metrics, the edge effects are highly suppressed. The Figures confirm that the graph-based models outperform the binning method at all test shower energies. It is also seen that the \neighbours and the \DGCNN model show similar performance.

\section{Resource requirements}
\label{sec:resource_requirements}

\begin{figure}[htbp]
    \centering
    \includegraphics[width=\figwidth\textwidth]{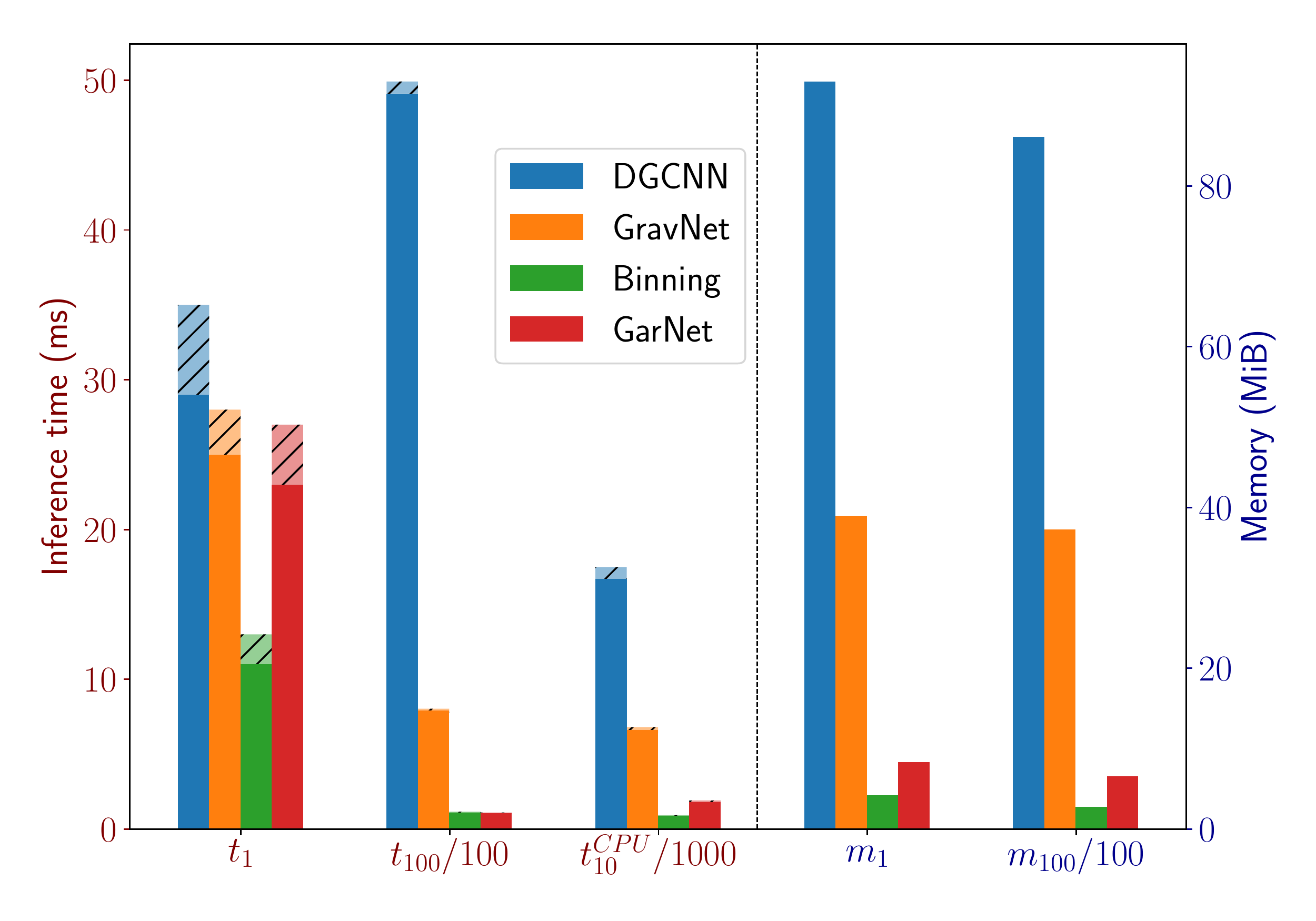}
    \caption{Comparison of inference time for the network architectures described in the text, evaluated on CPUs and GPUs with different choices of batch size. The shaded area represents the $+1\sigma$ statistical uncertainty band. \label{fig:timemem}}
\end{figure}

In addition to the clustering performance, it is important to take into account the computational resources demanded by each model during inference. The required inference time and memory consumption can have a significant impact on the applicability of the network for reconstruction tasks in constrained environments, such as the online and offline central-processing workflows of a typical collider-physics experiment. We evaluate the inference time $t$ and memory consumption $m$ for the models studied here on one NVIDIA GTX 1080 Ti GPU for batch sizes of 1 and 100, denoted as ($t_1, m_1$) and ($t_{100}$, $m_{100}$), respectively. The inference time is also evaluated on one Intel Xeon E5-2650 CPU core ($t_{10}^\text{CPU}$) for a fixed batch size of 10.
As shown in Fig.~\ref{fig:timemem}, memory consumption and execution times differ significantly between the models. The binning approach outperforms all other models, because of the highly optimized CNN implementations. The \DGCNN model requires the largest amount of memory, while the model using the \neighbours layers requires about 50\% less. The \aggregators model provides a good compromise of memory consumption with respect to performance. In terms of inference time, the binning model is the fastest and the graph-based models show a similar behaviour for small batch sizes on a GPU. The \aggregators and the \neighbours model benefit from parallelizing over a larger batch. In particular, the \aggregators model is mostly sequential, which also explains the outstanding performance on a single CPU core, with almost a factor of 10 shorter inference time compared to the \DGCNN model.

\section{Conclusions}
\label{sec:conclusions}

In this work, we introduced the \aggregators and \neighbours layers, which are distance-weighted graph networks capable of learning irregular patterns of sparse data, such as the detector hits in a particle physics detector with realistic geometry. Using as a benchmark problem the hit clustering in a highly granular calorimeter, we show how these network architectures offer a good compromise between clustering performance and computational resource needs, when compared to CNN-based and other graph-based networks. In the specific case considered here, the performance of the \aggregators and \neighbours models are comparable to the CNN and graph baselines. On the other hand, the simulated calorimeter in the benchmark study is only slightly irregular and can still be represented by an almost regular array. In more realistic applications, e.g. with the hexagonal sensors and the non-projective geometry of the future HGCAL detector of CMS, the difference in performance between the graph-based approaches and the CNN-based approaches is expected to increase further, making the \aggregators approach a very efficient candidate for fast and accurate inference and the \neighbours approach a good candidate for high-performance reconstruction with significantly less resource requirements but similar performance compared to the \DGCNN model for a similar number of free parameters. 

It should also be noted that the \aggregators and \neighbours architectures make no specific assumption on the structure of the underlying data, and thus can be employed for many other applications related to particle and event reconstruction, such as tracking and jet identification. Exploring the extent of usability of these architectures will be the focus of follow-up work.

\section*{Note added}
After the completion of this work, Ref.~\cite{Qu:2019gqs} appeared, discussing the application of a similar approach to the problem of jet tagging.

\section*{Acknowledgments}

We thank our CMS colleagues for many suggestions received in the development of this work. The training of the models was performed on the GPU clusters of the CERN TechLab and the CERN CMG group. This project has received funding from the European Research Council (ERC) under the European Union's Horizon 2020 research and innovation program (grant agreement n$^o$ 772369).


\bibliographystyle{ieeetr}

\begin{thebibliography}{10}

\bibitem{Denby:1987rk}
B.~H. Denby, ``{Neural Networks and Cellular Automata in Experimental
  High-energy Physics},'' {\em Comput. Phys. Commun.}, vol.~49, 1988.

\bibitem{Peterson:1988gs}
C.~Peterson, ``{Track Finding With Neural Networks},'' {\em Nucl. Instrum.
  Meth.}, vol.~A279, 1989.

\bibitem{Abreu:1992jp}
P.~Abreu {\em et~al.}, ``{Classification of the hadronic decays of the Z0 into
  b and c quark pairs using a neural network},'' {\em Phys. Lett.}, vol.~B295,
  1992.

\bibitem{Denby:1999kv}
B.~H. Denby, ``{Neural networks in high-energy physics: A ten year
  perspective},'' {\em Comput. Phys. Commun.}, vol.~119, 1999.

\bibitem{YANG2005370}
H.-J. Yang, B.~P. Roe, and J.~Zhu, ``{Studies of boosted decision trees for
  MiniBooNE particle identification},'' {\em Nucl. Instrum. Meth.}, vol.~A555,
  2005.

\bibitem{Radovic:2018dip}
A.~Radovic {\em et~al.}, ``{Machine learning at the energy and intensity
  frontiers of particle physics},'' {\em Nature}, vol.~560, no.~7716, 2018.

\bibitem{CMS-DP-2017-013}
V.~Khachatryan {\em et~al.}, ``{CMS Phase 1 heavy flavour identification
  performance and developments},'' Tech. Rep. CMS-DP-2017-013, 2017.

\bibitem{CMS-DP-2017-027}
V.~Khachatryan {\em et~al.}, ``{New Developments for Jet Substructure
  Reconstruction in CMS},'' Tech. Rep. CMS-DP-2017-027, 2017.

\bibitem{Pol:2018nhq}
A.~A. Pol {\em et~al.}, ``{Detector monitoring with artificial neural networks
  at the CMS experiment at the CERN Large Hadron Collider},'' {\em Comput.
  Softw. Big Sci.}, vol.~3, 2019.

\bibitem{krizhevsky2012imagenet}
A.~Krizhevsky, I.~Sutskever, and G.~E. Hinton, ``Imagenet classification with
  deep convolutional neural networks,'' {\em Commun. ACM}, vol.~60, pp.~84--90,
  May 2017.

\bibitem{HGCAL-TDR}
{CMS Collaboration}, ``{The Phase-2 Upgrade of the CMS Endcap Calorimeter},''
  Tech. Rep. CERN-LHCC-2017-023. CMS-TDR-019, 2017.

\bibitem{CMS-TP}
V.~Khachatryan {\em et~al.}, ``{Technical Proposal for the Phase-II Upgrade of
  the CMS Detector},'' Tech. Rep. CERN-LHCC-2015-010. LHCC-P-008.
  CMS-TDR-15-02, 2015.

\bibitem{carminati2017calorimetry}
F.~Carminati {\em et~al.}, ``Calorimetry with deep learning: particle
  classification, energy regression, and simulation for high-energy physics.''
  {"Deep Learning for Physical Sciences" workshop at NIPS 2017}, 2017.

\bibitem{guest2018deep}
D.~Guest, K.~Cranmer, and D.~Whiteson, ``{Deep Learning and its Application to
  LHC Physics},'' {\em Ann. Rev. Nucl. Part. Sci.}, vol.~68, 2018.

\bibitem{deOliveira:2018lqd}
L.~De~Oliveira, B.~Nachman, and M.~Paganini, ``{Electromagnetic Showers Beyond
  Shower Shapes}.'' {arXiv:1806.05667[hep-ex]}, 2018.

\bibitem{FCChh-CDR}
A.~Abada {\em et~al.}, ``Fcc-hh: The hadron collider,'' {\em The European
  Physical Journal Special Topics}, vol.~228, pp.~755--1107, Jul 2019.

\bibitem{cogan2015jet}
J.~Cogan {\em et~al.}, ``{Jet-Images: Computer Vision Inspired Techniques for
  Jet Tagging},'' {\em JHEP}, vol.~02, 2015.

\bibitem{komiske2017deep}
P.~T. Komiske, E.~M. Metodiev, and M.~D. Schwartz, ``{Deep learning in color:
  towards automated quark/gluon jet discrimination},'' {\em JHEP}, vol.~01,
  2017.

\bibitem{de2016jet}
L.~de~Oliveira {\em et~al.}, ``{Jet-images — deep learning edition},'' {\em
  JHEP}, vol.~07, p.~069, 2016.

\bibitem{baldi2016jet}
P.~Baldi {\em et~al.}, ``{Jet Substructure Classification in High-Energy
  Physics with Deep Neural Networks},'' {\em Phys. Rev.}, vol.~D93, no.~9,
  2016.

\bibitem{de2017learning}
L.~de~Oliveira, M.~Paganini, and B.~Nachman, ``{Learning Particle Physics by
  Example: Location-Aware Generative Adversarial Networks for Physics
  Synthesis},'' {\em Comput. Softw. Big Sci.}, vol.~1, no.~1, 2017.

\bibitem{paganini2018calogan}
M.~Paganini, L.~de~Oliveira, and B.~Nachman, ``{CaloGAN : Simulating 3D high
  energy particle showers in multilayer electromagnetic calorimeters with
  generative adversarial networks},'' {\em Phys. Rev.}, vol.~D97, no.~1, 2018.

\bibitem{rukhkhattak2018three}
G.~Rukhkhattak, S.~Vallecorsa, and F.~Carminati, ``Three dimensional energy
  parametrized generative adversarial networks for electromagnetic shower
  simulation,'' {\em 2018 25th IEEE International Conference on Image
  Processing (ICIP)}, pp.~3913--3917, 2018.

\bibitem{Musella:2018rdi}
P.~Musella and F.~Pandolfi, ``{Fast and Accurate Simulation of Particle
  Detectors Using Generative Adversarial Networks},'' {\em Comput. Softw. Big
  Sci.}, vol.~2, no.~1, 2018.

\bibitem{komiske2017pileup}
P.~T. Komiske {\em et~al.}, ``{Pileup Mitigation with Machine Learning
  (PUMML)},'' {\em JHEP}, vol.~12, 2017.

\bibitem{ATL-PHYS-PUB-2017-003}
{ATLAS Collaboration}, ``{Identification of Jets Containing $b$-Hadrons with
  Recurrent Neural Networks at the ATLAS Experiment},'' Tech. Rep.
  ATL-PHYS-PUB-2017-003, 2017.

\bibitem{Louppe:2017ipp}
G.~Louppe {\em et~al.}, ``{QCD-Aware Recursive Neural Networks for Jet
  Physics},'' {\em JHEP}, vol.~01, 2019.

\bibitem{Qu:2019gqs}
H.~Qu and L.~Gouskos, ``{ParticleNet: Jet Tagging via Particle Clouds}.''
  {arXiv:1902.08570[hep-ph]}, 2019.

\bibitem{Komiske:2018cqr}
P.~T. Komiske, E.~M. Metodiev, and J.~Thaler, ``{Energy Flow Networks: Deep
  Sets for Particle Jets},'' {\em JHEP}, vol.~01, 2019.

\bibitem{Nguyen:2018ugw}
T.~Q. Nguyen {\em et~al.}, ``{Topology classification with deep learning to
  improve real-time event selection at the LHC}.'' {arXiv:1807.00083[hep-ex]},
  2018.

\bibitem{Sirunyan:2017ulk}
A.~M. Sirunyan {\em et~al.}, ``{Particle-flow reconstruction and global event
  description with the CMS detector},'' {\em JINST}, vol.~12, no.~10, 2017.

\bibitem{Aaboud:2017aca}
M.~Aaboud {\em et~al.}, ``{Jet reconstruction and performance using particle
  flow with the ATLAS Detector},'' {\em Eur. Phys. J.}, vol.~C77, no.~7, 2017.

\bibitem{scarselli2009graph}
F.~Scarselli {\em et~al.}, ``The graph neural network model,'' {\em IEEE
  Transactions on Neural Networks}, vol.~20, no.~1, 2009.

\bibitem{battaglia2018relational}
P.~W. Battaglia {\em et~al.}, ``Relational inductive biases, deep learning, and
  graph networks.'' arXiv:1806.01261, 2018.

\bibitem{defferrard2016convolutional}
M.~Defferrard, X.~Bresson, and P.~Vandergheynst, ``Convolutional neural
  networks on graphs with fast localized spectral filtering,'' in {\em Advances
  in Neural Information Processing Systems 29} (D.~D. Lee, M.~Sugiyama, U.~V.
  Luxburg, I.~Guyon, and R.~Garnett, eds.), pp.~3844--3852, Curran Associates,
  Inc., 2016.

\bibitem{velickovic2017graph}
P.~Veli{\v{c}}kovi{\'{c}}, G.~Cucurull, A.~Casanova, A.~Romero, P.~Li{\`{o}},
  and Y.~Bengio, ``{Graph Attention Networks},'' {\em International Conference
  on Learning Representations}, 2018.

\bibitem{Selvi2018}
C.~Selvi and E.~Sivasankar, ``A novel adaptive genetic neural network (agnn)
  model for recommender systems using modified k-means clustering approach,''
  {\em Multimedia Tools and Applications}, vol.~78, pp.~14303--14330, Jun 2019.

\bibitem{Henrion2017NeuralMP}
I.~Henrion {\em et~al.}, ``Neural message passing for jet physics.'' {"Deep
  Learning for Physical Sciences" workshop at NIPS 2017}, 2017.

\bibitem{Abdughani:2018wrw}
M.~Abdughani {\em et~al.}, ``{Probing stop with graph neural network at the
  LHC}.'' 2018.

\bibitem{Martinez:2018fwc}
J.~Arjona~Martinez, O.~Cerri, M.~Pierini, M.~Spiropulu, and J.-R. Vlimant,
  ``{Pileup mitigation at the Large Hadron Collider with Graph Neural
  Networks},'' {\em Eur. Phys. J. Plus}, vol.~134, p.~333, 2019.
\newblock {arXiv:1807.07988 [hep-ph]}.

\bibitem{gilmer2017neural}
J.~Gilmer, S.~S. Schoenholz, P.~F. Riley, O.~Vinyals, and G.~E. Dahl, ``Neural
  message passing for quantum chemistry,'' in {\em Proceedings of the 34th
  International Conference on Machine Learning - Volume 70}, ICML'17,
  pp.~1263--1272, JMLR.org, 2017.

\bibitem{wang2018dynamic}
Y.~Wang {\em et~al.}, ``Dynamic graph cnn for learning on point clouds.''
  arXiv:1801.07829 [cs.CV], 2018.

\bibitem{tensorflow2015-whitepaper}
M.~Abadi {\em et~al.}, ``{TensorFlow}: Large-scale machine learning on
  heterogeneous systems,'' 2015.
\newblock Software available from tensorflow.org.

\bibitem{agostinelli2003geant4}
S.~Agostinelli {\em et~al.}, ``{GEANT4: A Simulation toolkit},'' {\em Nucl.
  Instrum. Meth.}, vol.~A506, 2003.

\bibitem{ioffe2015batch}
S.~Ioffe and C.~Szegedy, ``Batch normalization: Accelerating deep network
  training by reducing internal covariate shift,'' in {\em Proceedings of the
  32nd International Conference on Machine Learning, {ICML} 2015, Lille,
  France, 6-11 July 2015}, pp.~448--456, 2015.

\bibitem{kingma2014adam}
D.~P. Kingma and J.~Ba, ``Adam: {A} method for stochastic optimization,'' in
  {\em 3rd International Conference on Learning Representations, {ICLR} 2015,
  San Diego, CA, USA, May 7-9, 2015, Conference Track Proceedings}, 2015.

\bibitem{DBLP:journals/corr/abs-1708-07120}
L.~N. Smith and N.~Topin, ``Super-convergence: Very fast training of residual
  networks using large learning rates.'' arXiv:1708.07120 [cs.LG], 2017.

\end{thebibliography}

\end{document}